\documentclass{article}

\usepackage{arxiv}
\usepackage[utf8]{inputenc} 
\usepackage[T1]{fontenc}    
\usepackage{hyperref}       
\usepackage{url}            
\usepackage{booktabs}       
\usepackage{amsfonts}       
\usepackage{nicefrac}       
\usepackage{microtype}      
\usepackage{lipsum}		
\usepackage{graphicx}
\usepackage{doi}
\usepackage{verbatim}
\usepackage{authblk}
\usepackage{xcolor}
\usepackage{fancyvrb}
\usepackage{fvextra}
\title{Predicting Emergency Department Visits for Patients with Type II Diabetes}

\author{%
    \textbf{Javad M Alizadeh}\textsuperscript{1,\href{https://orcid.org/0000-0003-1421-8281}{\includegraphics[scale=0.06]{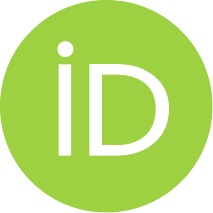}}}, 
    \textbf{Jay S Patel}\textsuperscript{2}, 
    \textbf{Gabriel Tajeu}\textsuperscript{1}, 
    \textbf{Yuzhou Chen}\textsuperscript{3}, \\
    \textbf{Ilene L Hollin}\textsuperscript{1}, 
    \textbf{Mukesh K Patel}\textsuperscript{1}, 
    \textbf{Junchao Fei}\textsuperscript{1}, 
    \textbf{Huanmei Wu}\textsuperscript{1, \href{https://orcid.org/0000-0003-0346-6044}{\includegraphics[scale=0.06]{orcid.pdf}}}\thanks{Corresponding author: \texttt{huanmei.wu@temple.edu}}
   }

 \affil{%
    \textsuperscript{1}Department of Health Services Administration and Policy, College of Public Health. \\
    \textsuperscript{2}Department of Oral Health Sciences, Kornberg School of Dentistry. \\
    \textsuperscript{3}Department of Computer and Information Sciences, College of Science and Technology. \\
    Temple University, Philadelphia, PA, USA.
}
\begin{document}
\maketitle

\begin{abstract}
	Over 30 million Americans are affected by Type II diabetes (T2D), a treatable condition with significant health risks. This study aims to develop and validate predictive models using machine learning (ML) techniques to estimate emergency department (ED) visits among patients with T2D. Data for these patients was obtained from the HealthShare Exchange (HSX), focusing on demographic details, diagnoses, and vital signs. Our sample contained 34,151 patients diagnosed with T2D which resulted in 703,065 visits overall between 2017 and 2021. A workflow integrated EMR data with SDoH for ML predictions. A total of 87 out of 2,555 features were selected for model construction. Various machine learning algorithms, including CatBoost, Ensemble Learning, K-nearest Neighbors (KNN), Support Vector Classification (SVC), Random Forest, and Extreme Gradient Boosting (XGBoost), were employed with tenfold cross-validation to predict whether a patient is at risk of an ED visit. The ROC curves for Random Forest, XGBoost, Ensemble Learning, CatBoost, KNN, and SVC, were 0.82, 0.82, 0.82, 0.81, 0.72, 0.68, respectively. Ensemble Learning and Random Forest models demonstrated superior predictive performance in terms of discrimination, calibration, and clinical applicability. These models are reliable tools for predicting risk of ED visits among patients with T2D. They can estimate future ED demand and assist clinicians in identifying critical factors associated with ED utilization, enabling early interventions to reduce such visits. The top five important features were age, the difference between visitation gaps, visitation gaps, R10 or abdominal and pelvic pain, and the Index of Concentration at the Extremes (ICE) for income.
\end{abstract}

\keywords{Type 2 Diabetes \and Machine Learning \and Prediction Model \and Emergency Department Visits}

\section{Introduction}

T2D is a prevalent chronic condition in the United States, impacting approximately 11\% of the adult population, which translates to about 37.3 million individuals \cite{1}. This disease is associated with a range of complications, including cardiovascular disease, chronic kidney disease, neuropathy, and impairments in vision and hearing, all of which significantly diminish quality of life, elevate morbidity rates, and contribute to mortality \cite{2, 3}. Remarkably, from 1981 to 2020, T2D consistently ranked among the top ten causes of death in the US \cite{4}. Moreover, the escalating healthcare expenditures, particularly concerning chronic illnesses, present a substantial financial burden on patients, insurance providers, and government budgets \cite{5,6,7,8}.
Emergency department (ED) visit rates per 100,000 individuals reached a peak in 2015 across all age groups, with a notable 10\% surge observed among individuals aged 45–64 from 2006 to 2015 \cite{9}. T2D is a prevalent chronic ailment among ED visitors, accounting for 12.8\% of ED visits in 2018 \cite{10}. Studies suggest that factors such as lower socioeconomic status, high social vulnerability, and limited access to primary care physicians contribute to heightened ED utilization \cite{11, 12} From 2008 to 2017, there was a substantial 55.6\% increase in all-cause diabetes-related ED visits per 10,000 adults, as reported by Uppal et al \cite{13}. Common clinical complications associated with these visits include uncontrolled blood sugar levels, abnormal blood pressure, irregular heart rates, and issues affecting the nervous system, kidneys, and eyes \cite{14}.

Fortunately, many ED visits related to T2D are preventable with a comprehensive understanding of patient-specific factors and social determinants of health (SDoH) contributing to these visits. By identifying and addressing these factors, clinicians, researchers, and policymakers can develop advanced preventive strategies. However, these factors are intricate and may vary based on geography, patient demographics, and socioeconomic conditions. Research underscores the importance of incorporating SDoH factors into health models due to their significant impact on population health \cite{15}. Nevertheless, studies utilizing electronic health records (EHR) often overlook SDoH factors. More systematic research is needed to discern which patient, community, and societal factors predict ED visits in diabetic patients.

Our objective is to develop multiple predictive models for ED visit risk, with a specific focus on patients with diabetes. We leverage data from a low socioeconomic area in greater Philadelphia, a demographic often underrepresented in research, to identify local risk factors critical for developing preventive measures and resource allocation. This study aims to: 1) establish a pipeline for preprocessing complex clinical data from various healthcare facilities, 2) integrate SDoH and EHR data, including visits, diagnoses, and vital signs, and 3) identify risk factors to construct multiple predictive models for ED visits among patients with diabetes. These identified risk factors will aid clinicians in prioritizing preventive strategies to ultimately reduce ED visits for diabetic patients.

\section{Materials and Methods}
\subsection{Data sources}

Data was extracted from the HSX Clinical Data Repository (CDR), a central database with encounter data and clinical details of patients' care journeys \cite{16}. We included adults aged 18 or older who received care at least once at participating HSX hospitals and had a diagnosis code for T2D in their electronic health records. Children with T2D and patients with Type I diabetes were excluded. In addition, hypertension is a common comorbidity among individuals with T2D. However, hypertension introduces additional variables that could potentially confound the study's results. For example, including patients with hypertension might lead to overfitting in the training process of ML models, as the majority of patients with T2D also have hypertension. For this study, we focused exclusively on patients with T2D but without hypertension. Further studies are working on patients with T2D and hypertension. By excluding this variable, we aim to reduce complexity, enhance the reliability of our findings, and isolate the effects and outcomes directly attributable to T2D. This approach allows for a clearer understanding of T2D progression and management in the absence of the confounding influence of hypertension. Additionally, this focus aligns with the study's objectives to explore specific outcomes or interventions relevant to T2D alone, ensuring that the findings are more targeted and actionable for this patient population. This decision is rooted in the need to provide precise and actionable insights that can directly inform clinical practice for patients with T2D, without the complexities introduced by the presence of hypertension.

The selected cohort consists of 34,151 unique patients with T2D but do not have hypertension, with at least one clinical encounter for primary care during the period from January 1, 2017, to December 31, 2022. The raw data retrieved includes 76,630,026 medical encounters, 113,897,301 vital signs, and 123,213,731 diagnoses, with detailed data variables for each table listed in Table 1. This dataset contains numerous duplicates due to system integrations and migrations.

Moreover, SDoH data for ZIP Code Tabulation Areas (ZCTA) is gathered from the American Community Survey and the Socioeconomic Data and Application Center (SEDAC) \cite{17, 18}. The Zip Code level SDoH data is associated with patients based on their provided Zip Code in patient demographics. Figure 1 shows an overview of the process.

\begin{table}[t] 
    \centering
    \caption{Raw data elements from HSX.}
    \includegraphics[width=0.7\linewidth]{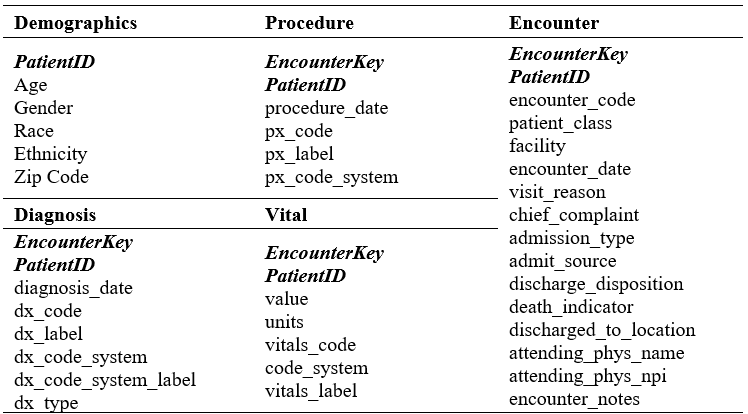}
    \label{tab:raw_data_elements}
\end{table}

\begin{figure}[t] 
    \centering
    \includegraphics[width=0.7\linewidth]{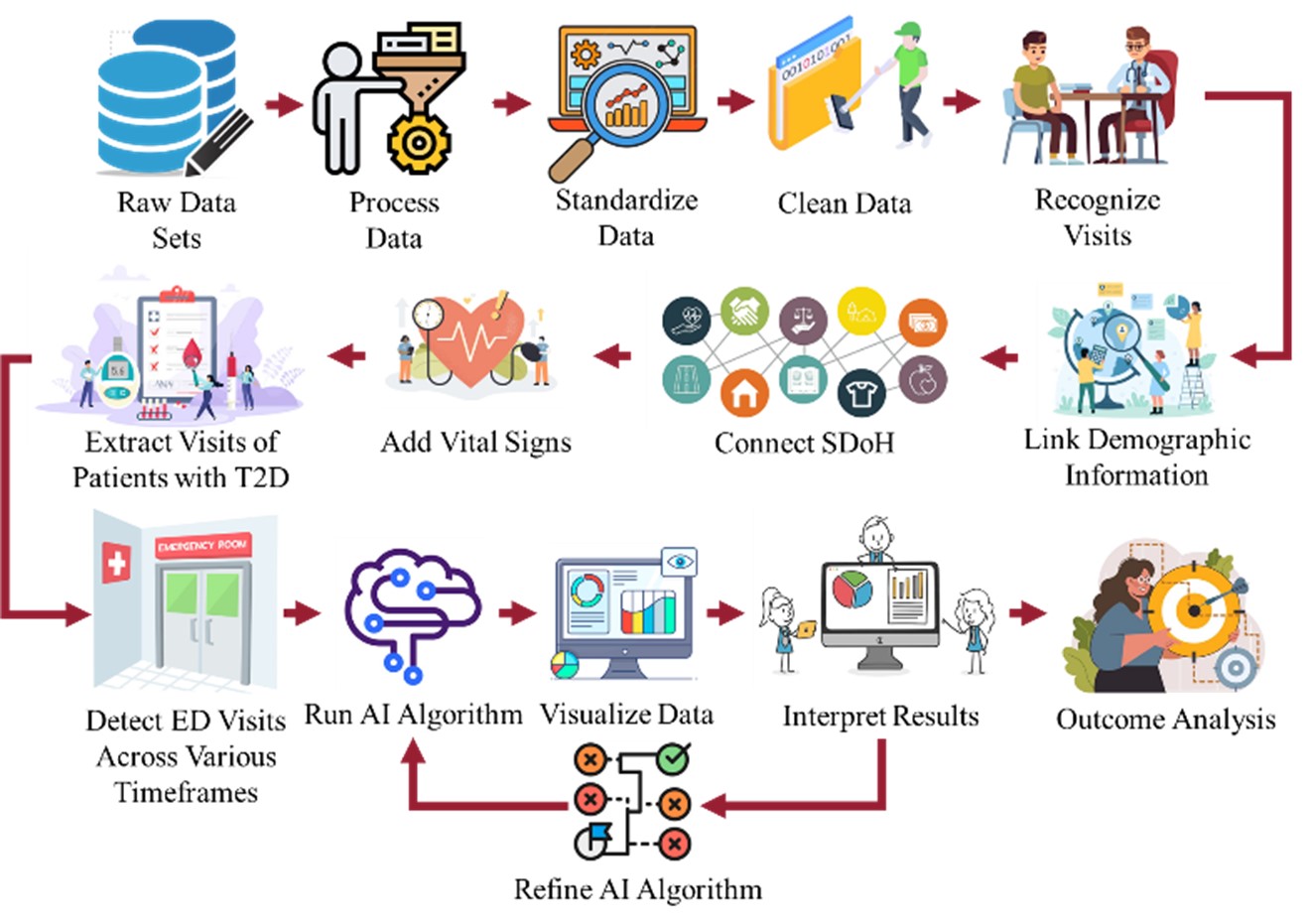}
    \caption{The overview of our project workflow.}
    \label{fig:project_workflow}
\end{figure}

\subsection{Data cleaning,  pre-processing, and standardization}

We utilized the following retrieved data tables from HSX: patient demographics, clinical encounters, vital signs, and diagnoses. We performed data quality assessments and descriptive statistics of the associated variables. Despite HSX’s implementation of real-time interoperability solutions, the data’s characteristics posed numerous challenges regarding quality. For example, vital sign measurements were in different units, such as “cm”, “inch” and “ft” for height, “kg”, “pounds” and “lb” for weight, and “Cel” and “[degF]” for temperature. In addition, there are different code systems for the diagnosis, such as ICD-9 and ICD-10. This heterogeneity requires standard conversions to ensure uniformity. For instance,  needing conversion to “inch” and “kg” to “pounds” or “lb”. Also, temperature units like “Cel” and “[degF]” require standard conversions to ensure uniformity.
Data standardization helps resolve data quality issues. Our team of health informatics experts and clinical professionals reviewed EHR systems, focusing on coding principles. By using Python library Pandas \cite{19}, we preprocessed the raw data for exploratory analysis through a data quality assurance pipeline:

\textit{Step 1. Normalize demographic information:}

Initially, the raw data had 8, 97, and 154 unique values for gender, ethnicity, and race, due to different coding systems, abbreviations, and acronyms. According to the Office of Management and Budget (OMB) Standards \cite{20}, we reorganized the race, ethnicity, and gender information based on Table 2 below. The age is normalized to years.

\begin{table}[t]
    \centering
    \caption{The standardized demographic information.}
    \includegraphics[width=0.7\linewidth]{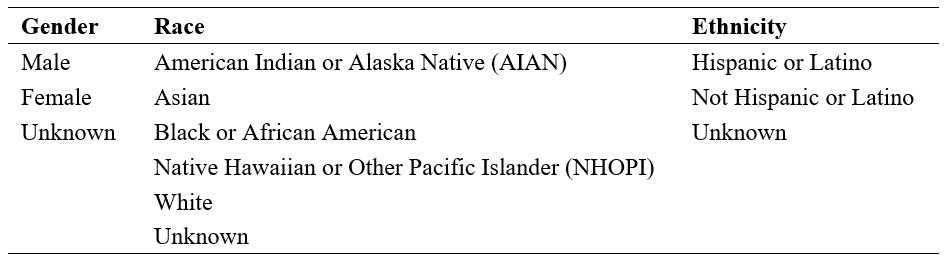}
    \label{tab:standardized_demographic_information}
\end{table}

\textit{Step 2. Mapping diagnosis codes to ICD-10:}

Seventeen diagnosis coding systems were used in raw data. The majority (70\%) is coded in ICD-10, followed by ICD-9 (8\%) and SNOMED-CT (2\%). We created a data dictionary, developed an algorithm, and implemented a software package using Python to map ICD-9 to ICD-10 codes. Efforts are ongoing to convert SNOMED-CT and others to ICD-10 using tools like the I-MAGIC reference application \cite{21}, but are omitted in the current dataset.

\textit{Step 3. Reduce diagnosis codes:}

The mapped ICD-10 diagnosis codes contained over 65,000 unique values. By extracting just the first three characters that represent the diagnosis category, we reduced this number to 742 distinct categories.

\textit{Step 4. Categorize patient class and admission type:}

We simplified "Encounter\_code," "Patient\_class," and "admission\_type" by combining similar groups and standardizing categories, according the suggestions of medical professionals. Uncomprehensible values were marked as “Missing”.

\textit{Step 5. Standardize vital signs:}

Vital units were standardized, ambiguities resolved, and "vitals\_label" entries unified. 

\textit{Step 6. Removing the provider information:}

Provider-related data, such as identifiers and names, were excluded to maintain patient privacy and focus solely on clinical and demographic insights.

\subsection{Connect SDoH factors with EMR}

Following data cleaning, the HSX dataset was filtered to include only patients with T2D. Each patient’s ZIP code was matched with the corresponding ZCTA and the associated SDoH indicators for that ZCTA. Consequently, for each patient, we now possess the following combined information:

\begin{itemize}
    \item The SDoH indicators for ZCTAs.
    \item Demographic information of patients obtained from EMR, including gender, race, ethnicity, and age.
    \item A compilation of comorbidities based on patient diagnosis codes, distinguishing between those with T2D and those without.
    \item Details of patient encounters, including newly calculated data such as first and last visit dates at HSX facilities, number of emergency visits, and the duration between an ED visit and the last encounter preceding it.
\end{itemize}

\subsection{Feature selection}

In our feature selection process for training machine learning models, we considered five categories: demographics (5 features), diagnoses (2,498 features), SDoH (30 features), vital signs (20 features), and computed features (2 features). We included all features from each category except for diagnoses, where we selected only the top 30 most frequent diagnoses across all visits. The rationale for this exclusion was that the remaining diagnoses had very low frequencies, rendering them negligible and potentially leading to biased outcomes due to insufficient representation. Additionally, we excluded the E11 code, which corresponds to T2D in the ICD-10 coding system, to mitigate the risk of overfitting during model training.

These categories were chosen because they encompass the most relevant and interpretable factors in understanding patient health, which directly impacts the predictive accuracy of models. By covering a broad spectrum of influences on health, these features allow models to make more comprehensive and accurate predictions, as they capture both clinical and non-clinical factors influencing health outcomes.

\subsection{Train and test machine learning models}

In this study, a variety of machine learning algorithms were employed to construct and evaluate predictive models. The applied algorithms include CatBoost \cite{22}, Ensemble Learning \cite{23}, K-nearest Neighbors \cite{24}, Support Vector Classification \cite{25}, Random Forest \cite{26}, and Extreme Gradient Boosting \cite{27}.

We converted the number of ED visits into a binary format to indicate whether a patient visited the ED. With over 2,000 distinct diagnoses, we selected the top 30 most frequent diagnoses and represented them in binary format. The dataset was divided into 70\% for training and 30\% for testing. To prevent overfitting and optimize hyperparameters, we used 10-fold cross-validation \cite{28}, training the model on nine folds and validating on one. Independent variables included demographics, SDoH, ED visits, common comorbidities, and vital signs. Machine learning algorithms were trained to predict ED visits based on these variables.

\section{Results}
\subsection{Data analysis}

In this study, we used 703,065 visits from a cohort of 34,151 patients diagnosed with T2D (T2D). Among these visits, 42,302 were to ED, accounting for 6\% of the total visits. Additionally, 14,799 patients, representing 43\% of the cohort, experienced at least one ED visit during the study period.

Figure 2 reveals significant variations in T2D prevalence across different ZIP codes. Particularly noteworthy are the exceptionally high percentages of T2D cases relative to population in ZIP codes 19132, 19133, and 19142. These areas stand out as having markedly elevated rates of T2D compared to surrounding regions.

Figure 3 illustrates the age distributions of patients with T2D segmented by gender and race, with the percentages representing the proportion of a gender/race group relative to the total population of that race group. For clarity, not all groups are shown. The data reveals notable age differences among patients with T2D across various racial and ethnic groups. Black patients tend to be older compared to their White counterparts, indicating a possible delay in the onset or diagnosis of T2D within the Black community. Additionally, there is a significant age difference between male and female patients with T2D, with a p-value of 0.001, indicating strong statistical significance. Among Black patients, the number of females with T2D is higher than that of males. This trend is particularly pronounced among Black females younger than 40 years of age, who exhibit the highest rate of diabetes across all racial groups.

\begin{figure}
    \centering
    \includegraphics[width=0.7\linewidth]{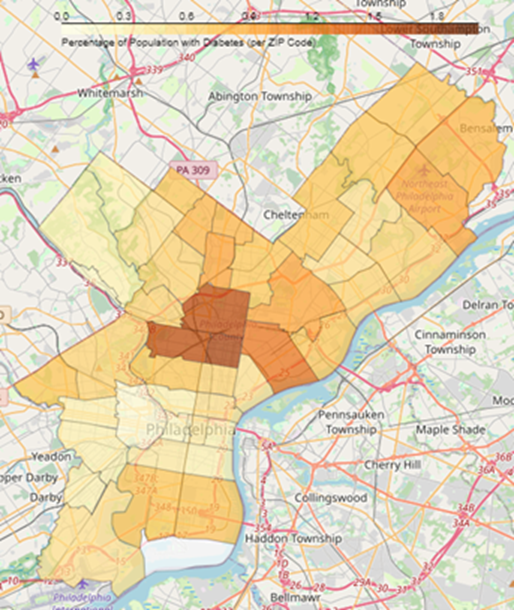}
    \caption{Percentage of T2D over ZIP code population.}
    \label{fig:percentage_T2D_over_ZIP_code_population}
\end{figure}

\begin{figure}
    \centering
    \includegraphics[width=1\linewidth]{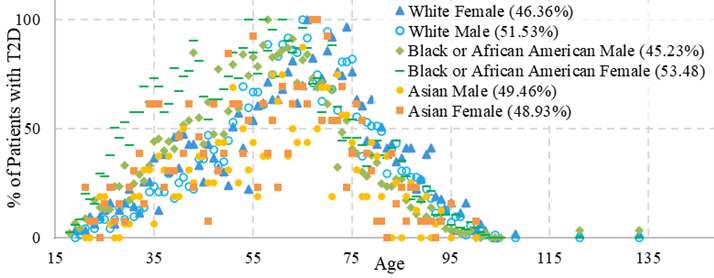}
    \caption{Age distributions of patients with T2D from Philadelphia among different gender and race groups. The percentage shows the proportion of a gender/race to the total population of a race group. Not all groups are shown for better illustration.}
    \label{fig:age-distributions}
\end{figure}

Figure 4 displays exploratory statistics on the gaps between visits for patients with T2D, with separate violin plots for ED and non-ED visits. The plots are divided into two subregions: (a) without outliers and (b) with outliers, providing a detailed view of the distribution and spread of visit intervals. For ED visits, the median gap is 22 days, indicating that half of the intervals between ED visits are less than 22 days. The first quartile (Q1) is at 6 days, suggesting that 25\% of the gaps are 6 days or less, while the third quartile (Q3) is at 71 days, indicating that 75\% of the visit gaps are 71 days or less. The minimum gap is 0 days, which means some patients have multiple ED visits on the same day. The maximum, excluding outliers, is 168 days, with an average gap of 64 days. This higher average compared to the median suggests the presence of longer intervals between some visits. When outliers are included, the plot shows a broader spread, capturing extreme values significantly beyond the 168-day mark, emphasizing the right skewness where a few patients have much longer gaps between ED visits. For non-ED visits, the median gap is 14 days, indicating that half of the intervals are less than this period. The first quartile (Q1) is at 5 days, and the third quartile (Q3) is at 42 days, showing that 75\% of the visit gaps are 42 days or less. The minimum gap is 0 days, indicating consecutive day visits for some patients, while the maximum, excluding outliers, is 98 days. The average gap is 43 days, which is slightly higher than the median, indicating moderate skewness. Including outliers extends the spread, particularly beyond the 98-day mark, showing similar right skewness with some significantly longer gaps.

\begin{figure}
    \centering
    \includegraphics[width=0.5\linewidth]{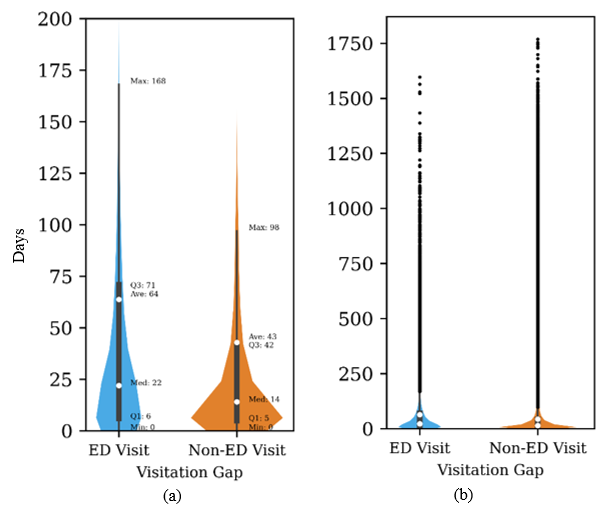}
    \caption{Exploratory visit statistics results of patients with T2D, the violin plots for gaps between ED and non-ED visits, (a) without outliers, and (b) with outliers}
    \label{fig:exploratory-visit-statistics}
\end{figure}

\subsection{Machine learning outcomes}

The performance of six machine learning models was evaluated on pilot datasets with 87 variables to predict ED visits in T2D patients, using a 70\%/30\% training-to-testing ratio. The models included Ensemble Learning (EL), Extreme Gradient Boosting (XGB), Random Forest (RF), CatBoost (CatB), K-nearest Neighbors (KNN), and Support Vector Classification (SVC). For fair comparison, all models are evaluated by using same metrics: AUC, accuracy, precision, recall, F1 score, and ROC. The results are summarized in Figure 5.

We observe that the EL, XGB, and RF each achieved an AUC of 0.82. EL had balanced metrics (accuracy, precision, recall, F1 score all at 0.74). XGB had slightly higher precision (0.75) but lower recall (0.72) and F1 (0.73). RF matched EL in all metrics. CatB had an AUC of 0.81, with accuracy, precision, recall, and F1 scores around 0.73.

\begin{figure}
    \centering
    \includegraphics[width=0.5\linewidth]{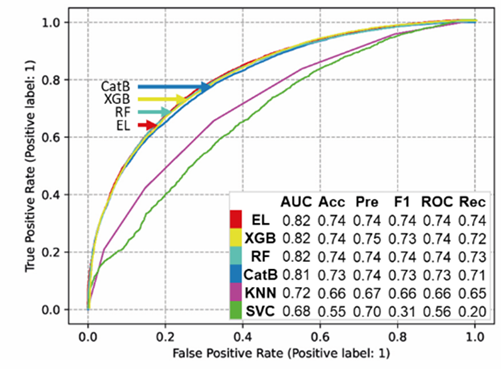}
    \caption{The Receiver Operator Characteristic (ROC) curves of the predictive models and their corresponding evaluation metrics.}
    \label{fig:roc}
\end{figure}

The CatB also yields a good performance but slightly worse than the EL, XGB, and RF. This indicates that it is slightly less effective than the top three models but still a competitive performer. The KNN showed a notable decline with an AUC of 0.72 and lower metrics, with accuracy, precision, recall, and F1 scores around 0.66 to 0.67, suggesting it is less suited for this task. The SVC had the weakest performance, with an AUC of 0.68, accuracy of 0.55, recall of 0.20, and F1 score of 0.31. Despite a precision of 0.70, the low recall indicates SVC frequently missed positive cases, resulting in poor overall predictive balance.

Figure 6 illustrates the feature importance from CatB, RF, and XGB, highlighting significant predictors of ED visits from demographic information, SDoH, diagnoses, and vital signs. Demographic features like age, race, and gender emerged as key predictors. Age is particularly crucial, as older individuals face higher risks for health conditions and T2D complications. Race and gender reflect health disparities and access to care, influencing ED visit frequency.

\begin{figure}
    \centering
    \includegraphics[width=0.5\linewidth]{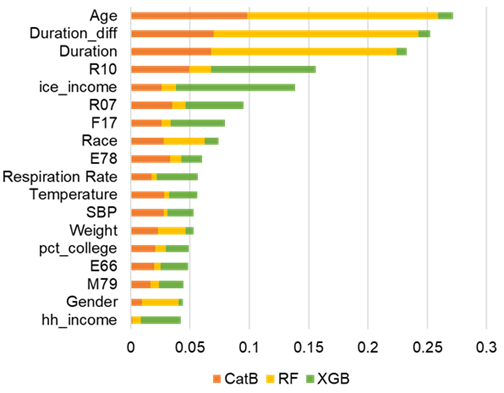}
    \caption{Features (represented by the ICD-10 codes, SDoH code, and vital signs) and the feature importance obtained from the three machine learning algorithms.}
    \label{fig:features-label}
\end{figure}

SDoH features such as income (ice\_income), education (pct\_college), and household income (hh\_income) were also important. These factors capture the socioeconomic context, significantly impacting health outcomes and access to healthcare. Higher income and education levels are typically associated with better health outcomes, underscoring their relevance in predicting ED visits among T2D patients.

Furthermore, several ICD-10 diagnostic codes were identified as important features, including R10 (abdominal and pelvic pain), R07 (throat and chest pain), F17 (nicotine dependence), E78 (lipoprotein metabolism disorders), E66 (obesity), and M79 (soft tissue disorders). These diagnoses are relevant due to their association with T2D and its complications. Vital signs like respiration rate, temperature, systolic blood pressure (SBP), and weight were also significant predictors, crucial for assessing patients' health status.

From our observations, auxiliary features such as the visitation gap (time between visits) and gap difference (difference in time gap between consecutive visits) showed importance, indicating healthcare utilization patterns. Figure 6 highlights a diverse range of features across demographic information, SDoH, diagnoses, vital signs, and auxiliary metrics as important predictors of ED visits among T2D patients, underscoring the multifaceted nature of T2D management. This comprehensive approach can guide targeted interventions and improve healthcare outcomes for T2D patients.

\section{Discussion }

In our preliminary data analysis of a diabetic patient cohort, we utilized SDoH and EMR data to develop a predictive model for ED visits using machine learning techniques. Our initial findings identified various risk factors that can guide targeted interventions aimed at reducing ED utilization and, consequently, healthcare expenditures over time. These results are consistent with previous studies and clinical observations reported in the literature. For instance, we found that conditions such as chest pain, upper respiratory tract infections, sprains and strains, and superficial injuries are among the top reasons for ED visits, aligning with established research on common ED utilization triggers \cite{29}.

Beyond the primary findings, our analysis uncovered several other SDoH factors, such as median home value, transportation access, and the social vulnerability index, that may influence ED visits. The association between some of these SDoH factors and ED utilization has not yet been systematically studied. These novel insights could be leveraged in clinical settings to recommend community resources that address these underlying issues. Additionally, our models identified occupation as a significant risk indicator. The ICE occupation indicator, which measures the polarization between white individuals in managerial or professional jobs and Black individuals in service, construction, and production industries, revealed that such polarized neighborhoods are highly predictive of ED visits. This highlights the importance of considering occupational and neighborhood disparities in understanding and mitigating ED utilization.

Moreover, the analysis of vital signs revealed their significant role in predicting ED visits among patients with T2D. Key indicators such as respiration rate, temperature, SBP, and weight emerged as critical features. These metrics are essential for assessing the current health status of patients and detecting acute changes that might necessitate emergency care. For instance, abnormal blood pressure and weight are directly linked to the management of T2D and its complications \cite{30}. The inclusion of vital signs in the predictive models underscores their importance in early identification of potential health deteriorations, allowing for timely interventions. By monitoring these vital signs, healthcare providers can better manage patients with T2D, potentially reducing the need for ED visits and improving overall health outcomes. This finding aligns with existing literature, which highlights the relevance of vital signs in chronic disease management \cite{31} and emergency care \cite{32}. Integrating these physiological measures into predictive analytics provides a more comprehensive approach to patient monitoring and risk stratification.

Furthermore, in real-world clinical settings, these factors hold relevance as patients often opt for ED visits due to severe pain when same-day appointments at their physician’s office are unavailable. Notably, smokers emerged as significant contributors to ED visits. Smoking is linked to various pulmonary conditions, cancers, and cardiovascular diseases (CVD), potentially prompting patients to seek ED care. Literature frequently highlights the heightened risk of CVD outcomes among diabetic individuals who smoke. Another noteworthy factor is “other soft tissue disorder,” encompassing conditions like Tendinitis, Bursitis, Myofascial pain syndrome, Rheumatoid Arthritis, and Lupus, which entail neurodegenerative diseases and often manifest in acute symptoms necessitating emergency care. While not directly diabetes-related, these conditions are prevalent among this population, likely due to the co-occurrence with higher body mass index. Clinicians should closely monitor complaints or diagnoses of other soft tissue disorders in diabetic patients to mitigate the likelihood of ED utilization. This underscores the practical utility of our model results in clinical practice.

Finally, to make the predictive interventions for managing ED visits by patients with T2D practical, a structured framework can be established. This framework would include the following components:

\begin{itemize}
    \item Capacity Planning and Resource Allocation

    \textit{Predict Future Demand:} Utilize the predictive model to forecast the volume of ED visits by T2D patients. This enables healthcare facilities to allocate resources effectively, ensuring that there are enough staff, beds, and specialized equipment available to manage the influx of patients. For instance, if the model predicts a spike in visits, additional insulin supplies and glucose monitoring equipment can be procured in advance.

    \textit{Real-time Monitoring:} Integrate the predictive model into the hospital’s EHR system to enable real-time monitoring of patient influx and resource utilization. This can trigger alerts for staff to prepare for increased demand during peak times.

    \item Targeted Interventions Based on Modifiable Features

    \textit{Risk Factor Identification:} Identify key features that contribute to higher risk of ED visits, such as poor control of blood glucose levels, high BMI, or infrequent healthcare visits. These modifiable risk factors can be the focus of targeted interventions.

    \textit{Patient Education and Recommendations:} Develop personalized care plans based on the identified risk factors. For instance, patients with high BMI could receive tailored advice on weight management, including diet and exercise recommendations. Patients with poorly controlled blood glucose levels could be advised to increase the frequency of their blood sugar monitoring and adjust their medication under healthcare provider supervision.

    \item Behavioral and Social Interventions

    \textit{Community Outreach and Support:} Engage with community health workers to provide education and support for T2D patients outside the hospital setting. This could include home visits, community workshops, or mobile health services that provide regular check-ups and education on managing diabetes.

    \textit{Patient Engagement and Self-management:} Encourage patients to take an active role in managing their condition by providing them with tools and resources, such as mobile apps for tracking blood sugar levels or reminders for medication adherence. This empowers patients to manage their health more effectively and potentially reduces their need for emergency care.

    \item System-level Integration and Feedback Loops

    \textit{Clinical Decision Support (CDS) Integration:} Embed the predictive model into the EHR system as a part of a CDS tool. This will allow healthcare providers to receive real-time alerts and recommendations when a patient’s data indicates a high risk of an ED visit, enabling preemptive action.

    \textit{Continuous Feedback and Model Updating:} Establish a feedback loop where data from actual patient outcomes is fed back into the model to continuously improve its accuracy. Regularly update the model with new data to reflect changes in patient behavior, treatment protocols, or other relevant factors.

    \item Policy and Institutional Support

    \textit{Multidisciplinary Collaboration:} Facilitate collaboration between different healthcare professionals (e.g., endocrinologists, dietitians, social workers) to provide comprehensive care for T2D patients. This team-based approach ensures that all aspects of the patient's health are addressed, reducing the likelihood of emergency visits.
    
    \textit{Institutional Policies:} Develop institutional policies that prioritize the integration of predictive models into routine care and allocate resources for the implementation and maintenance of these models.
    
\end{itemize}

\section{Acknowledgments}

The project is in part supported by a 2021 awarded grant under the Robert Wood Johnson Foundation’s Health Data for Action (HD4A) program, managed by AcademyHealth.

\end{document}